# The Concept of Cyber Defence Exercises (CDX): Planning, Execution, Evaluation

Ensar Seker
NATO CCD COE
Tallinn, Estonia
ensar.seker@ccdcoe.org

*Abstract* - **This paper discusses the concept of cyber defence exercises (CDX) that are very important tool when it comes to enhancing the safety awareness of cyberspace, testing an organization's ability to put up resistance and respond to different cyber events to establish the secure environment, gathering empirical data related to security, and looking at the practical training of experts on this subject. The exercises can give ideas to the decision makers about the precautions in the cybersecurity area and to the officials, institutions, organizations, and staff who are responsible on the cyber tools, techniques, and procedures that can be developed for this field. In the cyber defense exercises, the scenarios that are simulated closest to reality which provides very important contributions by bringing together the necessity of making the best decisions and management capabilities under the cyber crisis by handling stress and coordinated movement as a team. The objective of this paper is to address the issue from a scientific point of view by setting out the stages of planning, implementation, and evaluation of these exercises, taking into account and comparing international firefighting exercises. Another aim of the work is to be able to reveal the necessary processes that are required for all kind of cyber exercises, regardless of the type, although the processes involved vary according to the target mass of the planned exercise.**

## I. Related Work

CDX have been identified as an efficient mechanism to practice IT security awareness training [1, 2] but are also an ultimate tool to reveal and define the different security needs of every organization [3]. It provides an excellent opportunity and ultimate learning experience [4, 5] for the students to improve their skills in protecting and defending information systems are assessed in the context of realistic, true-to-life scenario [6]. On the other side, as discussed by Vigna [7] and Mink [8], the offensive security training is also an effective way to learn information security. The previous works in this area examined the structure [9] and how to use of cyber defence competitions, overall effectiveness of live-attack exercises in teaching information security [10], curriculum and course format at CDX in which teams design, implement, manage and defend a network of computers [11-15]. Other literature has examined the benefit of conducting cyber defense competitions at the K-12 level [16, 17]. The architecture of a cyber defense competition [18] and different tools and techniques used and how they fit into an active learning approach and how it focuses on the operational aspect of managing and protecting an existing network infrastructure were described by Green et al [19].

Patriciu and Furtuna [20] presented a number of steps and guidelines that should be followed when designing a cybersecurity exercise. One another approach of such live-attack exercises presented by White [21], lessons learned from illustrative examples of such exercises, as well as suggestions to help organizations conduct their own exercise. Other literature examined how to offer cyber defense competitions in the private sector, using a service provider model [22].

Existing literature has examined the potential benefits of cyber defense exercises. One another benefit of cyber defense exercise that can be instrumented to generate scientifically valuable modern labeled datasets for future security research [23, 24] and help uncover gaps in IT Security policies, plans and procedures [25]. It was claimed that [26] cyber exercises can be developed with a focus on measuring performance against specific standards. In cyber defense exercises, to measure team effectiveness and gain knowledge how to do that, the role of behavioral assessment techniques was investigated as a complement to task-based performance measurement [27].

In the literature, The RINSE simulator that is the real-time immersive network simulation environment for network security exercises was presented as a realistic rendering of network behavior [28]. In addition to that to execute real-time security exercises on a realistic inter-domain routing experiment platform was presented in the past [29]. A developed method for Job Performance Modeling (JPM) which uses vignettes for improving cybersecurity talent management through cyber defense competition design was described by Tobey [30].

## II. INTRODUCTION

Since the cyberspace was recognized as the fifth battlefield after land, sea, air, and space, it has begun to be of critical importance, especially in terms of national security. These attacks have become more popular in recent times, as cyber-attacks can be performed anonymously, declassified, and relatively inexpensive to perform in other areas. Therefore, countries have begun to use and develop cyberweapons at a sophisticated level of technology and sophisticated technology.

In order to safeguard the digital security of the public and the society, which is becoming an integral part of the national security, against cyber-attacks, states accept the idea that it is necessary to establish and disseminate authorities such as cyber defence commands, national cyber incident response teams, computer emergency readiness teams, and other information security centers. They also have begun to develop national cybersecurity strategies and put into practice.

In terms of cyber defense, cyber exercises have been playing a very important role in testing the technical cyber capacity of nations or organizations, cyber training, and cyber awareness raising that's why they have started to become widespread all over the world. Among the main objectives of the cyber defense exercises, they can increase [31, 32, 33, 34];

- The ability to test and develop common and coordinated technical and strategic mobility against the cyberattacks that may occur on a national basis.
- The Ability to test and develop common and coordinated technical and strategic mobility against cyber attacks, which may occur on an international basis.
- Ability to test and develop continuity and improving continuity processes with cybersecurity capabilities.
- Strengthening cooperation and coordination between public and private sectors in the cyberspace.
- Gathering empirical data related to cybersecurity research.
- The maturity level of legal and regulatory compliance.

In the following sections, the worldwide cyber defense exercises, and processes, types, and contributions of these exercises was examined with examples.

From the planning stage through to the implementation, execution and finally to the evaluation stage, cyber defense exercises can provide important contributions to both the exercise planners and their participants. These processes of exercises also can give an idea to a developer who develops mechanisms for the cyber defense.

## III. CYBER DEFENCE EXERCISES AROUND THE WORLD

Europe is at the forefront of the biggest players in the field of cyber defense exercises. In the past, 42 percent of the global cyber defense exercises have been carried out in Europe, as can be seen in Figure 1 [5].

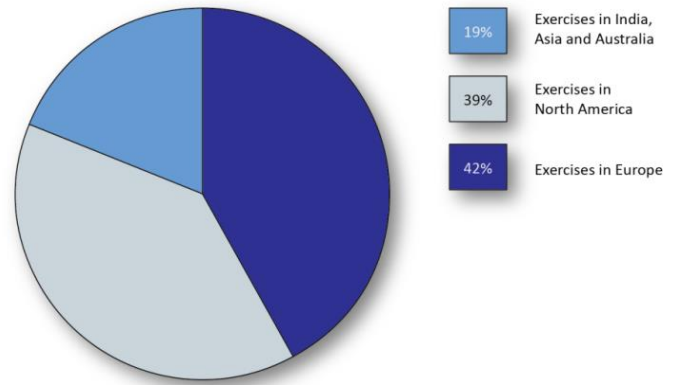

**Figure 1** - Cyber Defence Exercises (Worldwide Distribution)

Another actor, at least as important as Europe, in the field, is North America, especially the United States. North America is followed by Asia, mostly Japan, Malaysia, India, and Singapore. Australia comes right after Asia.

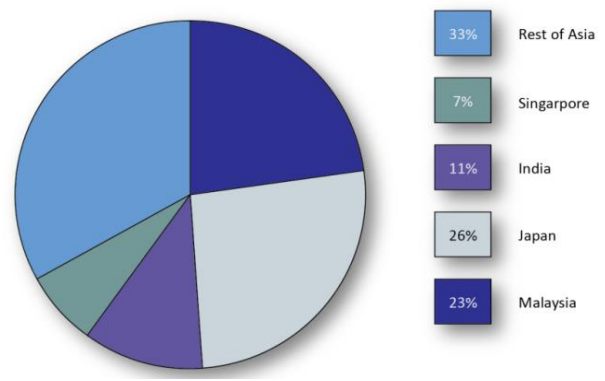

**Figure 2** - Cyber Defence Exercises (Asia Distribution)

Cyber defense exercises that were executed in Asia is as shown in Figure 2 [35]. It is noteworthy that Malaysia has moved up to second place just right after Japan, especially with investments that Malaysia has made in recent years to this field. Even though Japan's and Malaysia's number are close to each other, it is clear that Japan has much more experience in cyber defense exercises than in Malaysia.

Locked Shields, Cyber Coalition, Cyber Europe can be given as examples of a few cyber defense exercises carried out on an international scale.

*Locked Shields:* Locked Shields is organized annually by the NATO Cooperative Cyber Defence Centre of Excellence

(NATO CCD COE) in Tallinn, Estonia. It's accepted by the authorities that Locked Shields is the world's largest, most complex and technologically advanced cyber defense exercise. More than 900 cybersecurity experts from around the world have been involved in the 2017 Locked Shields exercise and the national teams of 20 countries have contributed. More than 2,500 attacks on the national teams (blue teams) were carried out by the red team in the exercise where more than 3,000 virtual systems were involved. Locked Shields exercises follow a successful route to adopting information technologies. As an example, the smart grid systems, air-to-air fueling systems, and drone control systems were added to exercise environment in 2017 [36].

*Cyber Coalition:* Cyber Coalition is organized annually by NATO. It's a three-day event and participation are from NATO members and alliance countries. In December 2016, more than 700 cyber defense and legal experts, government officials, officers, academics and industry representatives were involved in this exercise. In the exercise, there was also cyber defense personnel from the European Union and representatives of non-NATO countries such as Algeria, Austria, Finland, Ireland, Japan and Sweden [37].

*Cyber Europe:* It is organized by ENISA (European Union Agency for Network and Information Security), every two years for members of the European Union. Unlike military-based exercises such as Locked Shields and Cyber Coalition, It is organized by a civilian authority. The exercise, which took place in 2016, was included of the 28 European Union member states and 2 member countries of the European Free Trade Association (EFTA), although those countries are not a member of EU [38].

## IV. CYBER DEFENCE EXERCISES TAXONOMY

Cyber defense exercises can be carried out in various forms. According to the data set to be obtained, these types of exercises differ. However, the differences are based on parameters from ISO 22398, an international standard. It is possible to categorize cyber defense exercise into the following four groups for their objectives [39]:

i. Improvement and test of national/international cyber incident response cooperation.
ii. Evaluation/Competition of the cybersecurity skills, incident preparedness of individuals, organizations, and systems.
iii. Assessing information, readiness, capability, endurance and/or technical capacity.
iv. Training participants in real-world scenarios that provide the opportunity to gain knowledge, insight, experience by developing their skills and resilience before an incident.

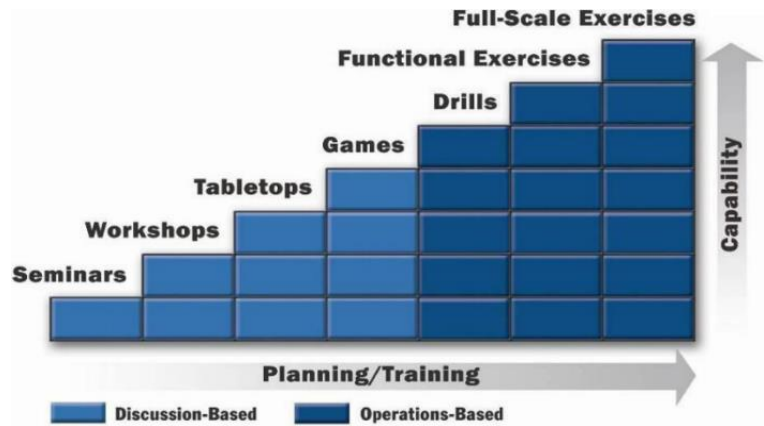

**Figure 3** – CDX by Type [40]

Despite all the different types of exercises that can be created such as capture the flag, discussion-based game, simulation, workshop, drills, seminar, cyber defense exercises can basically be divided into three categories [41];

1. Table Top Exercises: All scenarios / sub-scenarios, injections, and red team attacks are written and prepared before the exercise. In most cases, exercise planners and actors sit on a table and perform exercises, so such exercises are called table-top exercises. Table-top exercises should have a fairly limited number of training units and very well defined objectives [42]. The application process is relatively easy as it can be planned faster and in a shorter time than other types of exercises.
2. (Hybrid) Exercises: Exercise scenario / sub-scenarios and injections are pre-written, but the red team performs and attacks live during the exercise. Exercise planners perform exercises with a red team that applies real events according to pre-determined targets [43].
3. Full Live Exercises: In such exercises, while the main scenario and some sub-scenarios have already been prepared, there are instant scenarios and injections which are developed by the white teams (usually) according to the blue team's progress and strategies. The red team also develops new attack strategies based on the defensive capacity and status of the blue team. When compared with cyber defense exercises types, the planning process for full live exercises is much longer and more realistic [44].

## V. CYBER DEFENCE EXERCISES LIFE-CYCLE

In general, cyber defense exercises life-cycle has four major parts as follow [45];

*Identifying:* Includes topics such as recognizing and creating participants profile, determining the type and size of the exercise, evaluating current scenario options.

*Planning:* Includes topics such as Informing and training the people and teams involved in the exercise, setting up the media policy, inviting observers and media members, providing financial resources, setting the schedule and location of the exercise, distributing roles and creating a realistic scenario, preparing the exercise materials.

*Conducting:* Includes topics such as Implementation of the exercise in the most appropriate frame and rules, implementation of the scenarios and injections according to the determined sequence, resolution of the problems and faults that can occur during the exercise in the shortest and quickest manner, observation of participants and taking notes of decisions and activities of participants, and the management of the questionnaire and surveys for participants in order to support them.

*Evaluating:* Includes the creation of a group evaluating the exercises' results, the collection and evaluation of questionnaires and surveys answered by the participants, the collection of necessary information from the participants in the exercise, the preparation of documents to be submitted to the media, and the preparation of reports to be shared with the evaluators.

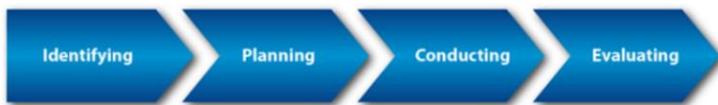

**Figure 4** - CDX Life Cycle [43]

## VI. PLANNING

### A. Determination of Objectives

A cyber exercise can be organized as an individual acting on an isolated network or as a broader training application on an operational network. Planning processes are similar. The exercise planning process begins with the determination of the purpose of the exercise and its desired outcome. Without explicit targets, planners cannot design a meaningful exercise. These goals allow participants to clearly configure scenarios in practice to determine whether they have the necessary skills in a cyber-environment against cyber threats. Different organizations have different guiding principles, tools, tactics, and procedures that make it important to create a starting point for each exercise.

- The determination of the effectiveness of the cyber training provided to participants before the exercise started,
- Evaluation of the effectiveness of exercise incident reports and preparation of analysis guides to address the deficiencies identified through the exercise,
- Evaluation of the ability of participants' necessary response against harmful activities and assess,
- Identifying the operational impacts of cyber-attacks and assessing the ability to implement recovery and recovery procedures required for these attacks,
- Determining the success of scenario planning and implementation,
- The elimination and correction of weaknesses in cybersecurity systems,
- Removal and correction of weaknesses in policies and procedures related to cyberspace,
- To maintain an information system and to determine what equipment or capabilities are required to carry out the necessary activities in a cyber-environment where harmful attacks are carried out,
- Determining whether the injections meet the objectives of the exercise,
- Increasing cyber awareness, readiness and coordination against cyber-attacks,
- Development of emergency preparedness plans for minimum damage, prevention by taking necessary precautions and protecting the information systems against cyber-attacks, are common targets set for all cyber defense exercises in general.

### B. Planning Process

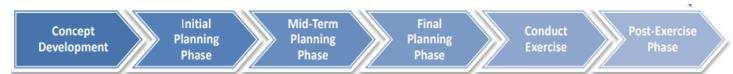

**Figure 5** - Planning Process [40]

*1) Initial Planning Meeting/Conference;*
Covers topics such as;
- Determination of requirements and conditions,
- Determination of scenario variables and draft scenario proposals,
- Collection of required information and distribution of tasks between execution planners,

It takes place approximately 6, 7 months before the exercises.

*II) Main Planning Meeting/Conference;*
Covers topics such as;
- Reconciliation of logistical and organizational problems such as personnel, scenario and time schedule development and administrative requirements,
- Examining, evaluating and finalizing all the draft documents to be used in the exercise,
- Examination and development of the injections prior to the final planning stage,

- Reviewing of the duties, conditions, and standards determined for the purpose of the exercise,

It takes place approximately 3, 4 months before the exercise.

### III) Final Planning Meeting/Conference;

The final planning conference is the final meeting to oversee the implementation processes and procedures. After this meeting, no major changes should be made to the design or coverage of the exercise or its supporting documentation. It takes place approximately 3, 4 weeks before the exercise.

### IV) Test-Run;

Test-Run is the final preparation phase for testing and evaluating the technical infrastructure of the cyber defense exercises. In test-run, it is aimed to establish all the sub-structures to be used for the exercise in the selected place for the exercise and to test these sub-structures as if the exercise process is working normally and to observe the possible problems before the exercise. Participation from all teams except the blue teams takes place in test-run. Thus, all teams will have the chance to review their final situation and practice process before the exercise. Test-run takes place approximately one week before the exercise.

## VII. EXECUTION

### A. Teams

#### I) Blue Team (Defensive);

Blue team is responsible for ensuring and defending the security of a company's or organization's information systems against virtual attackers (red team) in a virtual environment created within the scope of practice. In international cyber defense exercises, blue teams represent the national teams of each participating country. Against the simulated attacks, blue team should defend its network;
- over a given period of time,
- a defense based and operational context,
- following the exercise's rules.

Blue team also should identify and prevent any data leakage on their system The team also responsible for the protection of privacy, integrity, and usability of their network.

Since the cyber defense has been a part of national and international law and politics, media and national security strategies in recent times, the cyber defense exercises have also begun to be designed in this context. Only the technical defense by the blue team has begun to be seen as insufficient within the scope of cyber defenses. For this reason, legal, policy, strategy, and media scenarios have begun to be included in addition to technical scenarios, especially for international cyber defense exercises, so the responsibilities of the blue team have been increased.

The responsibilities of the Blue Team must always be observed within the framework of the rules of engagement [31, 32, 33, 34], the applicable laws and regulations, and any illegal action was taken by the team members is deemed unacceptable. Therefore, it is very important that all the actions and decisions taken by the blue team, even in the simulation environment, be performed without ignoring the existing laws and regulations.

Another clear rule in the rules of engagement is that the blue team cannot attack the exercise infrastructure, other blue teams, the red team and the virtual systems.

Blue team members must provide the right information, which will not harm their operational safety when requested.

Blue teams are able to communicate the green team that is responsible for exercises infrastructure, through the web page designed for them by submitting notifications and requests related to the technical problems about the exercise environment. The green team is responsible for resolving these technical problems within reasonable time.

It is important that all reports created by the team are made through the command chain within the team. Blue teams are allowed to use their own tools and software products, but all responsibility for the licensed copy of these products belongs to this team.

Within the white team, there is a group of people called 'blonde user' who are a response to occupy blue teams' users' services and systems. These users represent unconscious users and they may open harmful emails and files by clicking malicious links unconsciously. It is against the rules for the blue teams to deny these users' services and systems used by these users. It is also expected that the blue teams will be able to resolve the requests submitted by these users regarding technical problems related to the systems they are using, within a limited time.

In order to transfer preliminary information about the systems to be used with the exercise environment, blue teams are informed through webinars before the exercises.

#### II) Red Team (Offensive);

The aim of the red team is to achieve cyber-attacks equally to all the blue teams participating in the exercise. For this purpose, the red team follows a predefined scenario and has the permission to use security vulnerabilities that are already created in the blue team's systems. Successful attacks by the red team lead to a negative score for the blue teams. The red team and the white team must work closely together. The red team must always follow the instructions given by the white team. It is strictly forbidden for the red team to attack the services and infrastructure used by the green team. It is imperative that all attacks carried out by the Red Team remain within the exercise environment. This includes social engineering.

*III) Green Team;*

Green team is responsible for preparing and maintaining exercise systems and infrastructure. These infrastructures include systems that design, set up, and manage administrative computer nodes, virtualization platforms, storage, and core networking, as well as systems that blue teams must defend during the exercise. In order to ensure that these systems are functioning properly during the exercise, it is expected from the green team that they will be able to solve the technical problems submitted by the blue teams within a reasonable period of time.

*IV) Yellow Team;*

The role of the yellow team is to provide situational awareness during the exercise first for the white team and then for all participants in the exercise. The main sources of information for the yellow team are the interim reports provided by the blue teams, the reports of the attack campaigns from the red team members, and the reports provided by the system. The yellow team provides regular updates to white team leaders and blue teams.

*V) White Team;*

The white team is responsible for organizing the exercise and checking it during the execution. The white team determines the exercise objectives, the scenario, the high-level objectives for the red team, legal injections, rules, media preparations and communication plans. During the execution, the white team provides control of the exercise by determining when to start different stages, controlling the execution of the red team's campaign, and scoring issues. Management, blonde users, injections, scoring and media simulation are among the responsibilities of the white team.

*B. Scenario*

The desired outcomes of the exercise vary from one exercise to another, but these outputs always revolve around presenting realistic scenarios to demonstrate the cyber-threatening methods of participation and to evaluate the success of the exercise programs. Exercise outcomes should aim to raise awareness of various cyber threats and to give an idea to make a plan to prevent them. An example scenario of an international cyber defense exercise in the past years as follows; Country X is an island republic located in the western part of Africa and is a member of an international organization. There is a coalition force of this organization in the country. While the size of the island is comparable to that of Ireland, the climate and landscape are closer to Morocco. The Republic of X is a poor country, and especially sanitation, communication, medical services and education are quite inadequate. For example, the country has an insecure internet connection with the rest of the world, and the bandwidth of the connection is low. There are no law enforcement agencies or CERT to protect the country's information systems. This forces most international actors in the county to install and use expensive satellite communications or locally operated systems.

The Republic of X is in diplomatic conflicts with the Country of Y (a neighbor), which has been criticized by the international community for having a vigilant anti-democratic government. For a long time, the Republic of X is exposed to the cyber-attack, which is predicted to originate from the Country of Y. Immediately following the last diplomatic crisis between the Republic of X and the Country of Y, cyber-attacks started to take place at the Air Force base of the Republic of X and a number of confidential information and documents were stolen. As part of the international coalition, the mission of the blue team is to take necessary precautions at the Air Force base, analyzing IT devices, preventing ongoing and possible future attacks, and reporting to the HQ.

The Blue Team should try to fulfill the duties assigned to it in an unfamiliar system. They need to take in consideration the rules, media and strategy-based sub-scenarios and injections that will be included later throughout the exercise.

*C. Scoring*

Scoring is one of the most troubling issues for cyber defense exercises. Even if the scoring systems that are made are tried to be standardized, it is highly probable that objections always arise from the blue teams because scoring is usually made based on the initiative of the white team. For this reason, many cyber defense exercises, especially the ones that are organized at the international level, opposed the scoring system by arguing that scoring isn't the main purpose of the cyber defense exercises. Creating competition environment to build a better cyber defense is the main reason for score supporters. One of the examples of exercises that do not use the scoring system is Cyber Europe organized by ENISA. However, the use of the scoring system in such exercises was seen as a motivation tool for participants, and the positive competition between participants was a greater impetus for achieving more successful outcomes. Locked Shields, which was organized by NATO CCD COE, has been using scoring systems at the exercise.

*D. Monitoring*

Monitoring and logging is the basis for the scoring system and it helps identifying and responding to incidents during the exercise at an early stage. Cyber defence exercises are performed in a limited time and too many attacks and network activity occurs via exercise team members in this period that makes difficult for organizers to monitor corporate data being created across multiple networks and nodes. Therefore, monitoring provides a good understanding and in-depth analysis of fields in event logs and alerts created via Syslog, Nagios, DPI, NetFlow, etc. It is believed that controlling and scoring is one of the primary and critical asset in CDX that helps understanding real-time situation and performance of teams throughout the exercise and provides fine-grained control over network links and hosts.

*E. Media Activity Simulator*

The media simulator allows the actors to view and interact with the media and social media as if they were in real life. All players have their own passwords for social media use. Live to broadcast on all media and social platforms such as Twitter, Facebook, TV, radio, online news and newspapers are available through the simulator. With this simulation, web pages for the institutions and organizations of the host country are also available based on the scenario. While blue teams are busy taking the necessary precautions against attacks from red teams, they have to take the necessary steps in the media dimension as well like in real life.

*F. Injections*

Injections can be divided into 4 categories as; scenario injections, media games, legal games, and forensics.

1. scenario injections; scenario injections prepared by the white team that includes taking necessary precautions against cyber threat and vulnerabilities, following the news, evaluating intelligence, gathering information about cyber-attacks and preparing reports.

2. Media scenario; As mentioned before, the purpose of the media simulation is to bring the media environment to exercise environment so to challenge the blue teams even further. The stories in the news include information about events that occurred on ongoing cyber events, negative comments for the current cyber-attacks as well as fabricated news about them.

3. Legal games; The ability of the blue team to answer questions from the chain of command depends on having deep legal knowledge. To deal with complicated legal issues, to refute false statements and interpretations, and to communicate with the media in order to make the legal explanations related to the cyber-attacks intriguing to the general public understandable to those who are not experts and to respond in legal context to the news and analyzes that the media has published, are among the requirements of the game.

4. Forensic; Forensic aims to answer the questions related to current cyber-attacks such as who did it? what happened, when happened, how happened and why happened?

VIII. EXECUTION

One of the most important outputs of the cyber defense exercises is the After Action Report. At this report, it is mentioned the detailed performance of each blue team after the exercise. Addition to that the scenario and sub-scenarios, injections, exercise purposes, participants, scoring, technical infrastructure, red team attacks (client-side, web, network), defences made by the blue team, defects in these defences, general mistakes made, observations from all teams and sub-teams, recommendations and evaluations are also covered.

In addition, another private report is shared with each blue team separately about their specific performance team that includes analysis, evaluation, weaknesses and weak points, recommendations and recommendations during the practice.

IX. CONCLUSION AND FUTURE WORK

The importance of defense exercises is increasing day by day. It would be possible for countries to be involved in the global cyber defence exercises in the international arena, spreading the development and implementation of their own cyber defence exercise platforms on a national basis, and allocating higher budget figures to the planning and development of these exercises could contribute to achieving beneficial outcomes in the future to create stronger cyber defence systems. The emphasis on these exercises on the national and international scene will provide benefits in terms of uncovering the vulnerabilities in the area of the cyberspace, as well as the revitalization of the cyber defense awareness, and also the integrated technologies that can be followed in exercises related to the cyber defense.

As mentioned earlier for future studies, a technical tool will be developed on the scoring system, which is a problematic issue for cyber defense exercises, and on the standardization of this system and the development of a fairer system. As mentioned, the integration of new technologies such as power grid systems and drone control systems into cyber defense exercises is a critical issue. With the integration of these special systems into the exercises, the existing problems and the methods to be followed are another work to be done in the future.